\documentclass[prl,twocolumn,superscriptaddress,floatfix,amssymb,amsmath]{revtex4}

\usepackage{graphicx}
\usepackage{dcolumn}
\usepackage[latin1]{inputenc}
\usepackage[mathscr]{eucal}
\usepackage{epsfig}
\usepackage{rotating}
\usepackage{bm}

\newcommand{\TKK}{Department of Engineering Physics/COMP, Helsinki University of Technology P. O. Box
5100, 02015 TKK, Finland}
\newcommand{\LTL}{Low Temperature Laboratory, Helsinki University of Technology, P.O. Box 3500, 02015
TKK, Finland}

\begin{document}
\title{Experimental determination of the Berry phase in a superconducting charge pump}


\author{Mikko~M\"ott\"onen}\email{mikko.mottonen@tkk.fi}\affiliation{\LTL}\affiliation{\TKK}
\author{Juha J. Vartiainen}\affiliation{\LTL}
\author{Jukka P. Pekola}\affiliation{\LTL}

\begin{abstract} We present the first measurements of the Berry
phase in a superconducting Cooper pair pump. A fixed amount of
Berry phase is accumulated to the quantum mechanical ground state
in each adiabatic pumping cycle, which is determined by measuring
the charge passing through the device. The dynamic and geometric
phases are identified and measured quantitatively from their
different response when pumping in opposite directions. Our
observations, in particular the dependencies of the dynamic and
geometric effects on the superconducting phase bias across the
pump, agree with the basic theoretical model of coherent Cooper
pair pumping.
\end{abstract} \pacs{} 
\maketitle

\begin{figure}[h]
    \begin{center}
    \includegraphics[width=.45\textwidth]{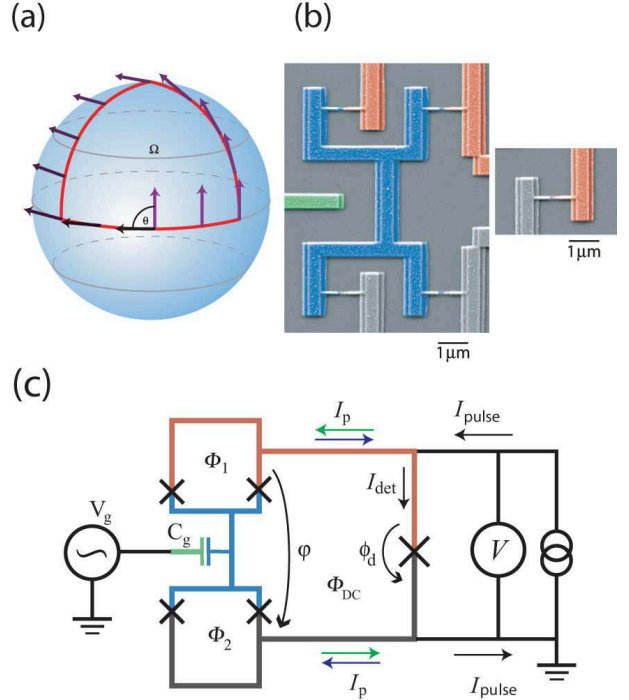}
    \end{center}
    \caption{\label{fig1}(a) Parallel transport of a vector along a path (red line) enclosing a solid
angle~$\Omega$. The lightest arrow shows the initial state of the vector and the darkest arrow the
final state. The resulting angle between the initial and the final vectors~$\theta$ is equal to the
solid angle~$\Omega$. (b) Scanning electron micrograph of the island on the left and of the detector
on the right. (c) Simplified circuit diagram of the measured sample. The corresponding parts in the
circuit diagram and SEM-images are marked by colors. The independent fluxes~$\Phi_1$, $\Phi_2$, and
$\Phi_{\rm DC}$ are controlled by on-chip coils~\cite{Vartiainen2007} and the gate voltage~$V_g$ is
related to the gate charge~$n_g$ and gate capacitance~$C_g$ by $n_g = C_gV_g/(2e)$.}
\end{figure}

Geometric phases arise from adiabatic cyclic evolution in
classical and quantum physics~\cite{Shapere1989}. In contrast to
dynamic effects, geometric phases depend only on the geometry of
the cycle traversed by the state of the system. In parallel
transport, for example, a vector is moved along a path without
changing its direction in a local coordinate
system~\cite{Shapere1989}. For a closed path in a flat Euclidean
space, the vector returns exactly to its initial state with
respect to a global coordinate system. In a curved space however,
the direction of the vector can change as shown in
Fig.~\ref{fig1}(a), where the vector moves on a surface of a
sphere along a loop enclosing a solid angle~$\Omega$. In this
case, the angle~$\theta$ between the initial and the final state
of the vector equals~$\Omega$ which depends only on the chosen
path, and hence is regarded as a geometric phase. In our universe,
geometric phases have been employed for example to measure the
curvature of space due to gravitation, and hence to test
Einstein's theory of general relativity~\cite{Ciufolini2004}. This
geodetic effect is intended to be measured as changes in the
rotation axes of gyroscopes inside a satellite orbiting Earth with
the stringent accuracy of $10^{-4}$ in the project Gravity Probe
B~\cite{Giulini2006}.

We consider adiabatic and cyclic temporal evolution in the ground
state of a quantum-mechanical system. The state of any pure
quantum system can be described by a complex valued wave function.
Thus the simplest geometric phase accumulated in a cycle, i.e.,
the Berry phase~\cite{Berry1984}, is a phase shift of the complex
number multiplying the wave function. As such, the absolute phase
of the wave function does not have a physical meaning, and hence
is unobservable. Thus the measurements of the Berry phase
typically rely on the interference of two states which have
undergone a different phase shift~\cite{Shapere1989, Anandan1997}.
This technique is also employed in the proposal to measure the
Berry phase in an asymmetric superconducting quantum interference
device (SQUID)~\cite{Falci2000}. In phase biased Cooper pair
pumps~\cite{Pekola1999} however, the accumulated Berry phase is
related to the pumped charge~\cite{Mottonen2006,Aunola2003}, and
hence we have a fundamentally different way to determine it. Here,
we report on the first experimental realization of phase biased
Cooper pair pumping in a superconducting loop.

Recently, superconducting circuits have proven to be suitable for
coherent manipulations of quantum
states~\cite{Makhlin2001,Vion2002}, in particular, two-level
quantum systems. The natural scalability of electric circuits
makes them potential candidates for qubits, i.e., basic building
blocks of the emerging quantum computer. On the other hand,
holonomies arising from adiabatic and cyclic evolution in a
degenerate eigenspace~\cite{Wilczek1984} are unitary
transformations which in turn, can be compiled to execute quantum
algorithms~\cite{Shor1997} of practical interest. In particular,
the holonomies related to charge transport in Cooper pair devices
have been studied theoretically~\cite{Brosco2007}. Thus the
observation of the Berry phase in superconducting circuits is an
important step towards the development of holonomic quantum
computation~\cite{Zanardi1999}. To date, holonomic quantum
computation has only been demonstrated using liquid state nuclear
magnetic resonance~\cite{Jones2000} (NMR), the scalability of
which is limited to about ten qubits.

\begin{figure}[h]
    \begin{center}
    \includegraphics[width=.45\textwidth]{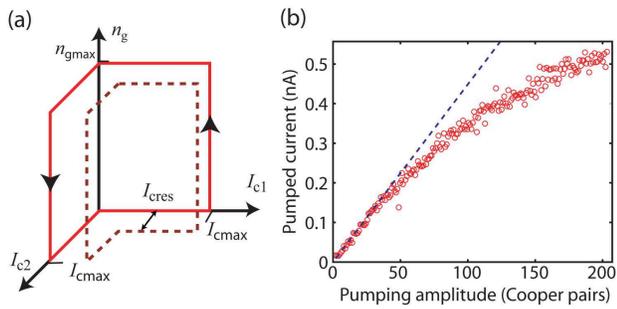}
    \end{center}
    \caption{\label{fig2}(a) Ideal pumping cycle (solid line) in the control parameter space of the
sluice. The tunable critical currents of the SQUIDs in the sluice $I_{c1} = I_{c1} (\Phi_1)$ and
$I_{c2} = I_{c2}(\Phi_2)$ are modulated together with the gate charge at the pumping frequency $f
=14\textrm{ MHz}$ for all the data in this paper.  In this cycle, the two critical currents are
analogous to valves of the charge flow, and the gate charge to a piston. The dashed line illustrates a
pumping cycle with finite~$I_{\rm cres}=\delta I_{\rm cmax}/2$ introducing finite dynamic current. (b)
Pumped current as a function of the maximum gate charge~$n_{g}^{\rm max}$. The blue line shows the
theoretical value obtained from Eq.~(\ref{eq1}) which is expected to be valid in the case of adiabatic
pumping for small pumping amplitudes $n_{g}^{\rm max}$. There are no fitted parameters since the
conversion constant of the gate voltage to gate charge is obtained from DC measurements.}
\end{figure}

Single-charge pumping is based on transporting a controlled
number~$n$ of carriers with quantized charge~$e^*$  in a cycle
repeated at frequency~$f$. This principle yields ideally the
pumped charge $Q_p=ne^*$, and hence the pumped current
\begin{equation}\label{eq1} I_p = ne^*f. \end{equation} The
carrier charge~$e^*$ is~$e$ for single electrons and~$2e$ for
Cooper pairs. In charge pumps described by a phase coherent order
parameter field, the phase difference of the field across the
device~$\varphi$ may play a significant role~\cite{Pekola1999}. In
fact for a constant~$\varphi$, adiabatic charge pumping gives rise
to the Berry phase~\cite{Mottonen2006,Aunola2003}. Interestingly,
the accumulated Berry phase, $\Theta_B$, in a cycle is related to
the pumped charge, $Q_p$, by \begin{equation}\label{eq2} Q_p =
-e^* \partial \Theta_B / \partial \varphi. \end{equation} In an
ideal pumping cycle corresponding to Eq.~(\ref{eq1}), the
accumulated Berry phase equals~$-n\varphi$. However, this kind of
pumping does not yield definite fingerprints on the relation
between the Berry phase and the pumped charge. Therefore, it is
important to reveal the phase coherent nature of pumping from its
dependence on~$\varphi$. Allowing a non-vanishing average
persistent current, i.e., leakage, through the pump during the
cycle, the pumped charge becomes phase dependent. In the regime
where charge states of the pump are approximately the eigenstates
of energy, one obtains in the two-charge-state approximation for a
cycle described in
Fig.~\ref{fig2}(a)~\cite{Pekola1999,Niskanen2003,Mottonen2006}
\begin{equation}\label{eq3} Q_p\approx e^* n(1-\delta\cos\varphi),
\end{equation} where $\delta\ll 1$ is proportional to the leakage
of the pump. Due to its geometric origin, the pumped charge is
independent of the pumping frequency. The charge leaked through
the pump during a cycle, i.e., the dynamic part of the
transferred charge is obtained as \begin{equation}\label{eq4}
Q_d\approx TI_\textrm{cmax}\delta\beta\sin\varphi, \end{equation}
where $T=1/f$ is the period of the pumping cycle,
$I_\textrm{cmax}$ is the maximum critical current in the pumping
cycle, and~$\beta$ is a constant specific to the particular pump
and to the control parameter cycle. To justify the validity of the
above model in our measurements, we compare the measured pumped
current $I_p=Q_pf$ and the dynamic current $I_d =Q_df$ with
Eqs.~(\ref{eq3}) and~(\ref{eq4}). Here, $I_d$ is the supercurrent
in the ground state of the system averaged over the pumping path
and thus it does not depend on~$f$ or~$n$, nor on the direction of
traversing the pumping path. On the contrary, $I_p$ is
proportional to both~$f$ and~$n$ in the adiabatic evolution, and
its sign changes on reversing the path. Agreement with the theory
allows us to determine the accumulated Berry phase from
Eqs.~(\ref{eq2}) and~(\ref{eq3}) as \begin{equation}\label{eq5}
 \Theta_B\approx-n(\varphi-\delta\sin\varphi).
\end{equation}

Figure~\ref{fig1}(b) and~(c) shows the Cooper pair pump, the
sluice in a 800~$\mu$m$^2$ superconducting loop with a detector
junction. In the ideal pumping cycle shown in Fig.~\ref{fig2}(a),
an integer number of~$n_{g}^{\rm max}$ excess electron pairs is
first attracted to the island through one SQUID and then repelled
from the island through the other one using the gate voltage and
the tunable critical currents of the SQUIDs in analogy with a
piston pump. Hence the sluice generates ideally an average pumped
current given by Eq.~(\ref{eq1}) with $n=n_{g}^{\rm max}$ The
details of the working principle of the sluice can be found in
Refs.~\cite{Niskanen2003,Mottonen2006}.

The Josephson junctions denoted by black crosses in the circuit
diagram of Fig.~\ref{fig1}(c) consist of AlO$_\textrm{x}$ tunnel
barriers fabricated by standard electron beam lithography and
two-angle evaporation into an all-aluminum device on oxidized
silicon wafer. The sluice part of our sample is identical to the
one used in Ref.~\cite{Vartiainen2007} except for up to 10\%
smaller junction size and stronger oxidation. The charging energy
of the island is difficult to measure in the presence of the
detector junction, and hence it is estimated based on the sample
used in Ref.~\cite{Vartiainen2007} to be 2~$\textrm{K}\times k_B$.
Using the Ambegaokar-Baratoff formula and the $IV$ characteristics
of the sample, the maximum critical currents of the SQUIDs and the
detector were estimated to be 30~nA each and 70~nA, respectively.
Due to the parallel structure of the junctions in the circuit,
there is an uncertainty of up to 20\% in the parameter estimation.
The plasma frequency of the detector is roughly 20~GHz.

We install a Josephson junction working as a threshold current
detector in parallel with the sluice forming a superconducting
loop as proposed in Ref.~\cite{Mottonen2006}. Figure~\ref{fig1}(c)
also shows our measurement scheme, in which we feed current pulses
through the circuit and monitor voltage across it. A voltage pulse
is observed if the system switches into the normal state in
response to the current pulse. The repetition rate of the
measurement was adjusted low enough for the switching events to be
uncorrelated. Such a detection method of circulating current was
realized, e.g., in the measurements of the superconducting qubit
Quantronium~\cite{Vion2002}.  To assure feasible operation, the
critical current of the detector is chosen to be much higher than
any instantaneous critical current of the sluice. The probability
for the system to switch to the normal state depends strongly on
the height of the external current pulse~$I_{\rm pulse}$. The
switching current~$I_{50}$ of the system is defined to be the
point where the probability is 50\%. We define backward pumping as
the direction for which the pumped current adds to the applied
400~$\mu$s current pulse, and hence the current through the
detector is given by $I_{\rm det}=I_{\rm pulse}+I_p-I_d$. For
forward pumping obtained by traversing the pumping cycle in the
opposite direction, the pumped current compensates part of the
external pulse: $I_{\rm det}=I_{\rm pulse}-I_p-I_d$. Thus we
measure shifts in~$I_{50}$ which correspond to twice the pumped
current. The average of the two switching currents equals to the
dynamic current plus a constant, namely, $I_{\rm det}$ at the
switching point. Note that we restrict our studies to take into
account only the dc component of the pumped current at the
detector. The effect of the ac component not filtered by the
circuit is left for future research.

Figure~\ref{fig2}(b) displays the measured pumped current as a
function of the gate amplitude for our most ideal pumping cycle.
For low enough gate amplitudes, the data shows a nearly linear
dependence in good correspondence with the theoretical behaviour
given in Eq.~(\ref{eq1}). For high gate amplitudes, the
adiabaticity of the pump breaks down and deviation from
Eq.~(\ref{eq1}) is observed. These results demonstrate the first
observation of Cooper pair pumping in closed superconducting
circuits~\cite{Mottonen2006,Rosenberg2006}. However, the data in
Fig.~\ref{fig2}(b) does not prove that the current arises from
coherent quantum dynamics.

The phase difference~$\varphi$ of the superconductor order
parameter across the sluice has to be a classical parameter for
Eq.~(\ref{eq2}) to hold. This is satisfied due to the detector
junction which protects the sluice from voltage fluctuations, and
hence phase biases it. For a detector junction with large enough
critical current, the phase difference~$\phi_d$ across it is
obtained from $\arcsin(I_{\rm det}/I_c)$ at the switching point
where $I_{\rm pulse} = I_{50}$. Thus we can control~$\varphi$ by
adjusting magnetic flux~$\Phi_{\rm DC}$ through the loop and
using the fundamental phase relation of a superconducting loop
$\varphi-\phi_d=2\pi\Phi_{\rm DC}/\Phi_0$, where $\Phi_0  =
h/(2e) = 2.07\textrm{ fWb}$ is the flux quantum. Due to proper
magnetic shielding, the flux offset in $\Phi_{\rm DC}$ was
negligible.

\begin{figure}[h]
    \begin{center}
    \includegraphics[width=.45\textwidth]{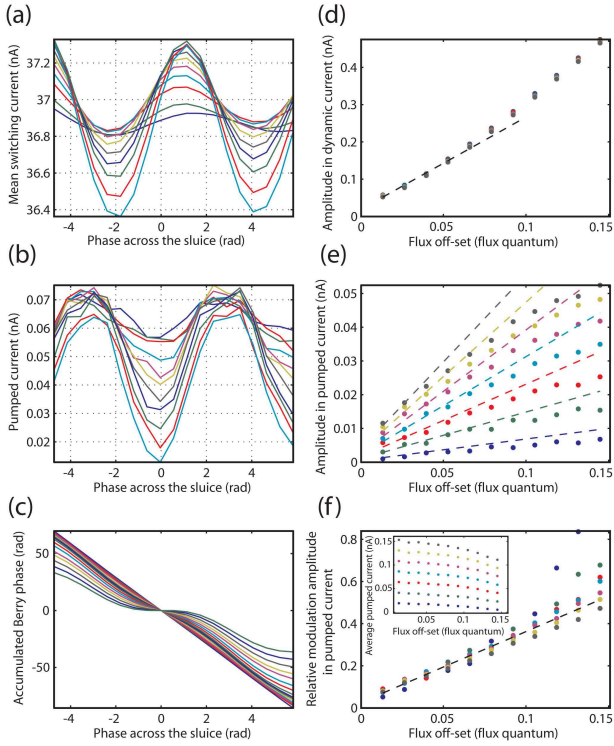}
    \end{center}
    \caption{\label{fig3}(a) Typical switching currents of the system averaged over forward and
backward pumping directions as a function of the phase~$\varphi$
across the sluice for several residuals of the critical
current~$I_{\rm cres}$, see Fig.~\ref{fig2}(a). Here, $n_{g}^{\rm
max}=16$. (b) Same as panel (a) but for the pumped current. To
obtain the amount of accumulated Berry phase, we make a fit to
the data corresponding to Eq.~(\ref{eq3}) with~$\delta$ and~$n$
as fitting parameters, the result of which is shown in panel~(c).
(d) The variation amplitude of the switching current as in
panel~(a) with respect to the phase across the sluice as a
function of the offset in the RF fluxes introducing residual
critical current for $n_{g}^{\rm max}=4$ (blue), 10 (green), 16
(red), 22 (cyan), 28 (magenta), 35 (yellow), and 41 (gray). These
values of the maximum gate charge are in the linear adiabatic
regime as shown in Fig.~\ref{fig2}(b). (e) The variation
amplitude of the pumped current as in panel~(b) from the same
data set as in panel~(d). (f) Modulation amplitude in the pumped
current divided by the average pumped current (shown in the
inset), $\delta$, as a function of the flux off-set. The dashed
line shows a linear fit to the data for eight smallest off-sets,
and hence yields the linear dependence of the leakage parameter
$\delta$ on the flux off-set. The dashed line in panel~(d) shows
a corresponding theoretical line from Eq.~(\ref{eq4}) using
$\beta=0.0245$ and $I_\textrm{cmax}=30$~nA. The dashed lines in
panel~(e) show the linear dependence of $\delta$ multiplied by
the average pumped current (see the inset) corresponding to the
specific gate amplitude and the smallest flux off-set.}
\end{figure}

The evidence of the phase coherence in our measurement is shown in
Fig.~\ref{fig3}. Figure~\ref{fig3}(a) presents the variation of
the switching current of the system with respect to the phase
difference across the sluice controlled by the eternal
flux~$\Phi_{\rm DC}$ as described above, and Fig.~\ref{fig3}(b)
displays the pumped current from the same measurement. Clear
sinusoidal modulation is observed in both curve sets in agreement
with Eqs.~(\ref{eq3}) and~(\ref{eq4}), implying that the sluice
is coherent and phase biased. Note that the minimum of the pumped
current corresponds quite accurately to the point of vanishing
phase difference across the sluice and the dynamic current is
phase shifted by almost $\pi/2$ radians as in Eqs.~(\ref{eq3})
and~(\ref{eq4}). The pumped number of Cooper pairs~$n$ and the
relative modulation amplitude~$\delta$ are determined from the
data and the corresponding curves for the accumulated Berry
phases are shown in Fig.~\ref{fig3}(c) according to
Eq.~(\ref{eq5}). For the largest values of~$\delta$, second order
corrections to Eq.~(\ref{eq5}) may modify the estimated value for
the Berry phase.

To further test our scheme, we measured the dependence of the
modulation amplitudes on the gate amplitude and residual critical
current $I_\textrm{cres}=\delta I_\textrm{cmax}/2$, i.e., leakage,
shown in Figs.~\ref{fig3}(d) and~\ref{fig3}(e). The residual
critical current was introduced by an offset in the control fluxes
of the SQUIDs from the ideal pumping cycle. Figure~\ref{fig3}(f)
shows $\delta$ as a function of the offset. As predicted, the
modulation amplitude of both, the pumped and the dynamic current,
increase with the residual critical current, but only the pumped
current depends on the maximum gate charge~$n_{g}^{\rm max}$, in
agreement with the theory. We note that this way of measuring the
Berry phase is fundamentally different from the conventional
method utilizing interference with excited states.

Our observations pave the way for further experiments on Cooper
pair pumping in closed circuits~\cite{Rosenberg2006}, on the
quantum standard of electric current, for applications of
geometric phases in holonomic quantum
computation~\cite{Zanardi1999}, and test the fundamental
implications of the quantum theory in an order parameter
describing coherent dynamics of a macroscopic number of condensed
Cooper pairs. Our measurements are in agreement with a theoretical
model neglecting decoherence.~\cite{EPASP} . On the other hand, the effects of
dephasing and dissipation in superconducting qubits are harmful in
the manipulation of superconducting quantum systems, see, e.g.,
Refs.~\cite{Yoshihara2006}. Thus our results support the
robustness of geometric phases against decoherence and
fluctuations~\cite{Fazio2003}. We thank the Academy of Finland,
M.M. the Finnish Cultural foundation, Magnus Ehrnrooth Foundation,
and V\"ais\"al\"a foundation for financial support. R.\ Fazio, F.\
Hekking, D.\ Cohen, J.\ Ankerhold, P.\ Zanardi, J.\ Peltonen, A.\
Niskanen, M.\ Paalanen, and M.\ Nakahara are acknowledged for
discussions and A.\ Kemppinen, M.\ Meschke, O.-P.\ Saira, and A.\
Savin for technical support.

---\emph{Note added.} During the peer review of this work Leek et
al.\ reported on the observation of the Berry phase in a
superconducting qubit~\cite{Leek2007}.


\end{document}